# Multiple decoherence-free states in multi-spin systems

H. J. Hogben[1,*], P. J. Hore[1] and Ilya Kuprov[2]

[1]*Department of Chemistry, University of Oxford, Physical and Theoretical Chemistry Laboratory, South Parks Road, Oxford, OX1 3QZ, UK.*

[2]*Oxford e-Research Centre, University of Oxford, 7 Keble Road, Oxford OX1 3QG, UK.*

Email: hannah.hogben@chem.ox.ac.uk

**ABSTRACT**

A numerical procedure is presented for mapping the vicinity of the null-space of the spin relaxation superoperator. The states populating this space, *i.e.* those with near-zero eigenvalues, of which the two-spin singlet is a well-studied example, are long-lived compared to the conventional $T_1$ and $T_2$ spin-relaxation times. The analysis of larger spin systems described herein reveals the presence of a significant number of other slowly relaxing states. A study of coupling topologies for *n*-spin systems ($4 \leq n \leq 8$) suggests the symmetry requirements for maximising the number of long-lived states.

**Keywords**

NMR, singlet, decoherence, relaxation, long-lived state

**Introduction**

An ability to tame spin relaxation by finding the rare states that resist the gradual drift towards thermal equilibrium [1-3] is of considerable interest in magnetic resonance. The slowly relaxing states may be used to store polarization [4-7], measure slow diffusion [8], and obtain information on molecular structure [9]. Such super-stable states were recently found to exist in dipolar-coupled spin systems [1] and have since attracted much theoretical [7; 10-12] and experimental [2; 7-8; 11] attention. Their common property – a zero dipolar relaxation rate – can be formulated as defining the null space of the relaxation superoperator to which their associated eigenvectors belong. More generally, we are interested in finding those eigenstates of the full relaxation superoperator that have small (but not necessarily zero) eigenvalues. Analytical treatments are possible for small systems [3; 10-12], but collections of more than four spins have not hitherto been studied systematically.

Drawing on our recent work on large-scale spin dynamics simulations [13-15] and construction of relaxation superoperators [16] for large spin systems, we report here a systematic mapping of the dipolar relaxation null spaces for systems with four to eight spins using the *Spinach* software library [17]. Investigation of different spin-½ coupling topologies evinces clear symmetry and coupling requirements for maximising the number of long-lived states.

The null-space analysis follows a simple procedure: build the Liouvillian and find the eigenvectors corresponding to zero or near-zero eigenvalues [12; 18]. Different dipolar coupling topologies comprising at least four spins are predicted to have more than one non-trivial (*i.e.* other than the identity operator) long-lived state when certain symmetry requirements are met. These criteria are illustrated with three examples of commercially available molecules expected to exhibit multiple long-lived states in liquid state NMR experiments.

**Theoretical formalism**

In order to be long-lived, a state $\hat{\rho}$ or a subspace of states $K = \mathrm{span}\{\hat{\rho}_1,...,\hat{\rho}_k\}$ must be invariant with respect to time evolution under the system Liouvillian $\hat{\hat{L}}$:

$$\exp\left(-i\hat{\hat{L}}t\right)\hat{\rho} = \hat{\rho} \quad \Rightarrow \quad \left(\sum_{n=0}^{\infty}\frac{(-it)^n}{n!}\hat{\hat{L}}^n\right)\hat{\rho} = \hat{\rho} \quad \Rightarrow$$

$$\hat{\rho}+\sum_{n=1}^{\infty}\frac{(-it)^n}{n!}\hat{\hat{L}}^n\hat{\rho} = \hat{\rho} \quad \Rightarrow \quad \sum_{n=1}^{\infty}\frac{(-it)^n}{n!}\hat{\hat{L}}^n\hat{\rho} = 0 \quad \forall t \in [0,\infty)$$

(1)

where $\hat{\hat{L}} = \hat{\hat{H}} + i\hat{\hat{R}}$, $\hat{\hat{H}}$ is the Hamiltonian commutation superoperator and $\hat{\hat{R}}$ is the relaxation superoperator. This invariance is only possible (as a consequence of the Taylor expansion uniqueness theorem) if $\hat{\rho}$ or $K$ belongs to the null space of the Liouvillian:

$$\hat{\hat{L}}\rho = 0 \qquad \hat{\hat{L}}: K \to \{0\} \tag{2}$$

The general task of finding long-lived states in a particular system therefore amounts to finding the null space of the system Liouvillian. Control of the coherent part of the Liouvillian (contained in $\hat{\hat{H}}$) is well developed: field shuttling, spin-locking and decoupling can be used to halt the evolution of states of interest under the influence of chemical shift differences or external *J*-couplings [11-12;19]. There remains the part of the Liouvillian that we cannot normally control – the incoherent relaxation processes – which has yet to be fully exploited from a practical perspective. Accordingly, we shall focus our attention on the relaxation superoperator and look for states that are immune to it. The most general formulation of spin-relaxation theory for spin systems with a one-way coupling to a classical heat bath is [20-21]:

$$\hat{\sigma}(t+\Delta t) = \hat{\sigma}(t) + \left[\sum_{n=1}^{\infty}(-i)^n \int_0^{\Delta t}dt_1 \int_0^{t_1}dt_2 \ldots \int_0^{t_{n-1}}dt_n \left\langle \hat{\hat{H}}_1^R(t_1)\hat{\hat{H}}_1^R(t_2)\ldots\hat{\hat{H}}_1^R(t_n)\right\rangle\right]\hat{\sigma}(t) \tag{3}$$

where $\langle\ \rangle$ denotes ensemble average, $\hat{\sigma}(t)$ is the ensemble average density matrix and $\hat{\hat{H}}_1^R(t)$ is the stochastic part of the Liouvillian commutation superoperator, both written in the interaction representation with respect to the coherent Liouvillian. Truncation of this series at the second order with some additional assumptions is known in magnetic resonance as Bloch-Redfield-Wangsness relaxation theory [22-23]. As most semi-classical theories do, this formalism relaxes the system to the infinite temperature state. This is easily fixed [24], but the relaxation destination is unimportant for the present treatment – we only seek to determine whether a particular state is going (or not going) to relax at all.

It is clear from Equation (3) that a *sufficient* condition for a state to be immune to propagation under the relaxation superoperator is belonging to the null space of the stochastic Hamiltonian commutation superoperator at all times:

$$\hat{\hat{H}}_1^R(t)\hat{\sigma} = 0 \quad \Rightarrow \quad e^{i\hat{\hat{H}}_0 t}\hat{\hat{H}}_1(t)e^{-i\hat{\hat{H}}_0 t}e^{i\hat{\hat{H}}_0 t}\hat{\rho} = 0 \quad \Rightarrow \quad \hat{\hat{H}}_1(t)\hat{\rho} = 0. \tag{4}$$

Whether or not this is also a *necessary* condition, we do not know – this seems unlikely, because terms could in principle cancel or vanish under the average in Equation (3). This explains why singlet states are immune to dipolar relaxation (the singlet density matrix commutes with the dipolar Hamiltonian involving the two spins in question and with any operator that does not affect those spins) and provides a rapid test for the long-lived character of a state of interest. In particular any state from which all transitions under $\hat{\hat{H}}_1(t)$ are permutation-symmetry-forbidden is going to be long-lived. However, *all* of the long-lived states (including those where the eigenvalue is very small but non-zero) can only be discovered with certainty by mapping the null-space of the full relaxation superoperator in square brackets in Equation (3).

Before embarking on larger systems, we shall briefly review the classic two-spin case [1]. For a dipolar-coupled pair of spins-½ the null space of the relaxation superoperator is spanned by two states. If only dipolar relaxation is considered, the eigenvalues are exactly zero and the states persist indefinitely. Any linearly independent pair of states (they do not have to be orthogonal) may be chosen as the basis of this two-dimensional null space, for example:

$$\hat{1} = \hat{P}_{T_+} + \hat{P}_{T_0} + \hat{P}_{T_-} + \hat{P}_S \quad \text{and} \quad \hat{P}_{TS} = \hat{P}_{T_+} + \hat{P}_{T_0} + \hat{P}_{T_-} - \hat{P}_S \tag{5}$$

where $\{\hat{P}_S, \hat{P}_{T_+}, \hat{P}_{T_0}, \hat{P}_{T_-}\}$ are projectors into the singlet and the three triplet states. The identity operator $\hat{1}$ is not interesting as because it commutes with any Hamiltonian and will hereafter be omitted from the discussion. The $\hat{P}_{TS}$ state is the difference between the singlet state, $\hat{P}_S$, and the sum of the three triplet states. Any linear combination of these states is also in the null space and is thus also long-lived. A particularly convenient (on symmetry grounds) linear combination, known as "the" singlet state may be written in the usual $\alpha\beta$ basis set as:

$$\hat{P}_S = \tfrac{1}{2}(|\alpha\beta\rangle - |\beta\alpha\rangle)(\langle\alpha\beta| - \langle\beta\alpha|), \tag{6}$$

corresponding to a projector into the two-spin zero-quantum coherence. For convenience, the normalisation factors will henceforth be omitted.

The magnetic dipolar interaction operator, which often dominates relaxation in liquid-state NMR, is bilinear in the spin operators:

$$\hat{H}_D = D(r)\left[3(\hat{\boldsymbol{S}}_1 \cdot \boldsymbol{u}_{12})(\hat{\boldsymbol{S}}_2 \cdot \boldsymbol{u}_{12}) - (\hat{\boldsymbol{S}}_1 \cdot \hat{\boldsymbol{S}}_2)\right] \tag{7}$$

with $D(r) = -\mu_0 \hbar \gamma_1 \gamma_2 / 4\pi r_{12}^3$. $\hat{\boldsymbol{S}}_i$ is a vector of the spin operators, $\{\hat{S}_x, \hat{S}_y, \hat{S}_z\}$, $r_{12}$ is the distance between the spins and $\boldsymbol{u}_{12}$ is the unit vector along the spin-spin direction. In large spin systems, because products and linear combinations of two-spin singlets are invariant under sums of dipolar operators, we would expect some of the long-lived states to be products and linear combinations of products of two-spin singlets.

Our large-scale implementation of liquid-state Bloch-Redfield-Wangsness relaxation theory is described in detail elsewhere [16]; it includes all cross-correlations and non-secular terms as well as (optionally) dynamic frequency shifts. The isotropic rotational diffusion approximation is used in this paper, but the Spinach library [17] can handle arbitrary user-specified correlation functions. If chemical shielding anisotropy (CSA) is present, it is also included because singlet states are not immune to CSA and (DD)-CSA processes [1; 19].

In the simulations presented below the dipolar interactions were calculated directly from user-specified coordinates (model systems) and crystal structure geometries (real molecules). CSA tensors and scalar couplings were estimated using the DFT GIAO B3LYP/EPR-II method in Gaussian03 [25]. All interaction tensors were then fed into the relaxation theory module of the Spinach library [17], the resulting relaxation superoperator was diagonalized and the eigenvectors corresponding to small eigenvalues were inspected. An example of MATLAB code calling the *Spinach* library for this purpose can be found in the Supplementary Information along with the console output that lists all the interaction parameters.

**Results and Discussion**

A systematic search to find long-lived states by mapping the null space of the full relaxation superoperator has been performed on dipolar coupled systems with three to eight spin-½ nuclei. A range of system symmetries and coupling patterns was investigated; the results are summarised in Fig. 1 in which coupling topologies are grouped according to the number of long-lived states found in them. A nearest neighbour separation of 1 Å was chosen (to give a dipolar coupling of $D \approx 120$ kHz); being just shorter than a $^{13}$C–H bond length this provides an upper bound on the dipolar interactions likely to be encountered in organic molecules. All pair-wise dipolar interactions were included in the relaxation superoperator with coupling constants and dipolar axes calculated from the geometry of the spin system.

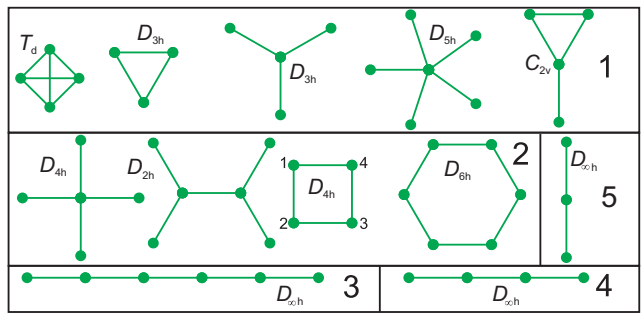

FIG 1. Dipolar coupled spin-systems grouped according to null-space size. Nodes represent spin positions and solid lines a separation of 1 Å ($D \approx 120$ kHz).

For a dipolar coupled three-spin-system only the linear geometry returns a long-lived state as was shown analytically in Ref. [26]. For larger systems it appears that only the topologies that have a centre of inversion show any long-lived states at all. This is to be expected for the dipolar relaxation superoperator, which inherits the symmetry of Equation (7) with respect to the permutation of the two spins. The long-lived states are not formed from combinations of singlets across the most strongly coupled pairs, *i.e.* those with minimum separation, but instead from the symmetric combinations of products of singlets across the spins interchanged by the inversion symmetry. For example, the square $D_{4h}$ four-spin system yields a non-trivial null-space state:

$$4\hat{P}_S^{\{1,3\}}\hat{P}_S^{\{2,4\}} - \mathbb{1} \tag{8}$$

(the numbering is shown in Fig 1). Because all linear combinations of eigenvectors with small eigenvalues do themselves have small eigenvalues, it is possible to take an alternative linear combination of the state in Equation (8) with its identity operator partner to yield the more intuitively recognisable $\hat{P}_S^{\{1,3\}}\hat{P}_S^{\{2,4\}}$. This demonstrates the localised singlet character of the inversion-related spins, *i.e.* of those across the diagonal of the square which experience the weaker of the two dipolar coupling amplitudes in the system. The same state, with its triplet mixture counterpart, also forms a part of the null space of the linear four-spin system (Fig. 1), along with two further mixed states. An isolated singlet state across any pair of spins in that system does not appear anywhere in the null space and thus would not exhibit slow relaxation. This may be confirmed by substitution of the requisite states into Equation (4) which fails to return the all-zero vector indicative of long-lived character.

Multi-spin systems are known to have a long-lived state when a localised singlet interacts only weakly with neighbouring spins [3]. Coupling topologies combining strong ($D \approx 120$ kHz) and weak ($D \approx 120$ Hz) interactions were therefore investigated with the results shown in Fig. 2 and Supplementary Information Table SI.

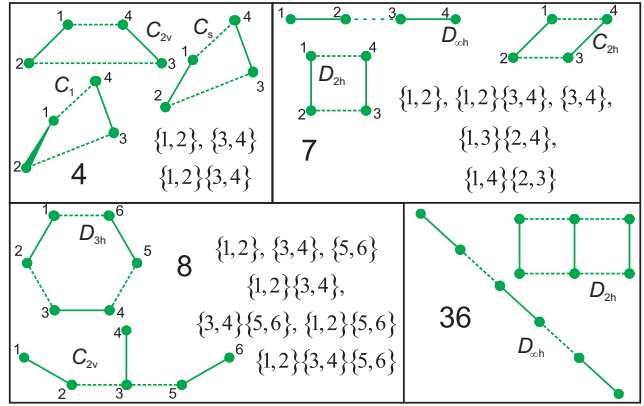

FIG 2. Multi-spin systems with both strong ($D \approx 120$ kHz, solid lines) and weak ($D \approx 120$ Hz, dashed lines) dipolar couplings, grouped according to null-space size, *n*. When $n < 10$ the constituent states of each group which derive from singlet-states, $\hat{P}_S^{\{i,j\}}$, across pairs of spins $\{i,j\}$ are given. Nodes represent spin positions. All pair-wise dipolar interactions were included.

Four-spin systems, of $C_1$ and $D_{2h}$ symmetry, have been observed experimentally to display a long-lived state [27]. The nature of the observed state was proposed to be a singlet localised on one of the two pairs of strongly coupled spins stabilised by the *J*- coupling within the pair [18]. The results from our analysis, shown in Fig. 2, demonstrate that this is certainly possible but that a more complicated product of singlets may also be responsible for the long lifetime observed. The states in addition to the isolated singlets are either products of the strongly dipolar-coupled singlets or symmetric linear combinations across all pair-wise singlet permutations. The number of long-lived states available is maximised when sets of strongly coupled pairs are related by centres of inversion, more precisely by an inversion with respect to the midpoint of the pairs, not necessarily the centre of inversion of the whole system. The state vectors are linear combinations or products of localised singlets across all pairs of symmetry related spins, even if they have only a weak dipolar coupling. As the interactions between sets of strongly coupled pairs increases the long-lived states are increasingly perturbed and the non-symmetric linear combinations mix so as to increase the eigenvalues of the relaxation matrix, reducing the long-lived states to those shown in Fig. 1.

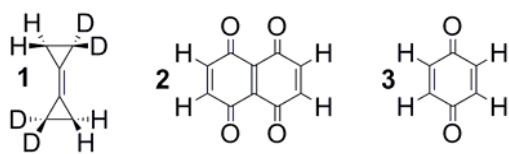

FIG 3. Molecules expected to show long lived states.

Having established the symmetry and coupling pattern requirements for the presence of multiple long-lived states, we have investigated three apparently suitable molecules, Fig. 3. They were chosen primarily for the symmetry of their dipolar coupling networks but also for their relatively small CSA. The eigenvalues of the system Liouvillian were calculated for three cases: dipolar relaxation superoperator only, full relaxation superoperator (including CSA) only, and finally the total Liouvillian that includes also the coherent evolution for the molecule in a moderate (1 Tesla) static field. Table 1 shows the 10 smallest amplitude eigenvalues in each instance; long-lived states are defined as those that have eigenvalues more than an order of magnitude smaller than the majority.

**Table 1.** Ten smallest eigenvalues (in s$^{-1}$) for dipolar relaxation only, dipolar and CSA relaxation and the full coherent Liouvillian for three example molecules, numbering as in Fig. 3, in a 1.0 T magnetic field. The identically zero eigenvalue always belongs to the identity operator. All simulation parameters for these systems are given in the supporting information. The magnetic properties were calculated using Gaussian03 [25] using literature crystal structures [28-30] .

| Molecule | Dipolar relaxation | Dipolar and CSA relaxation | Total Liouvillian |
|---|---|---|---|
| 1 | 0 | 0 | 0 |
|   | 0.0000 | 0.0009 | 0.0684 |
|   | 0.0015 | 0.0024 | 0.1271 |
|   | 0.0038 | 0.0046 | 0.7623 |
|   | 0.0057 | 0.0064 | 2.3863 |
|   | 0.0057 | 0.0065 | 2.8843 |
|   | 0.0087 | 0.0093 | 3.4775 |
|   | 1.8800 | 1.8694 | 3.9168 |
|   | 1.8800 | 1.8695 | 3.9168 |
|   | 1.8800 | 1.8810 | 5.6113 |
| 2 | 0 | 0 | 0 |
|   | 0.0000 | 0.0000 | 0.0000 |
|   | 0.0001 | 0.0003 | 0.0643 |
|   | 0.0003 | 0.0003 | 0.0644 |
|   | 0.0004 | 0.0004 | 0.1581 |
|   | 0.0004 | 0.0004 | 0.2967 |
|   | 0.0006 | 0.0006 | 0.2988 |
|   | 0.1710 | 0.1683 | 0.3294 |
|   | 0.1710 | 0.1683 | 0.3778 |
|   | 0.1710 | 0.1683 | 0.4549 |
| 3 | 0 | 0 | 0 |
|   | 0.0000 | 0.0000 | 0.0000 |
|   | 0.0033 | 0.0032 | 0.0648 |
|   | 0.0038 | 0.0038 | 0.0946 |
|   | 0.0085 | 0.0085 | 0.1612 |
|   | 0.0085 | 0.0085 | 0.3264 |
|   | 0.0088 | 0.0088 | 0.3382 |
|   | 0.1644 | 0.1666 | 0.3904 |
|   | 0.1644 | 0.1666 | 0.4158 |
|   | 0.1669 | 0.1669 | 0.5289 |

Each molecule reveals six non-trivial long-lived states, that is, six states that would resist relaxation if coherent evolution is suppressed. In the presence of unsuppressed coherent evolution, bicyclopropylidene-$d_4$ (**1**) naphthalenetetrone (**2**) and *p*-benzoquinone (**3**) all have only one extremely long-lived state, which has an eigenvalue seven orders of magnitude smaller than the average. Reducing the field strength, or using other methods to limit coherent evolution, moves the eigenvalues towards the relaxation-only limit, that is, the six usable states identified in Fig 2. Molecules with more sets of inversion-related dipolar coupled pairs should exhibit even more long-lived states.

**Conclusion**

We have presented a method with which to determine the number and identity of the states in multi-spin systems that are immune or insensitive to dipolar relaxation. Using this procedure we have identified some general requirements to maximise the number of long-lived states. Firstly, for a strongly coupled spin system to exhibit long-lived states, some spin pairs must be related by an inversion symmetry. Multiple pairs of strongly dipolar-coupled spins with only weak inter-pair interactions lead to many long-lived states when there are 'pairs of pairs' related to each other by a centre of inversion. We have identified large slowly relaxing subspaces in several commercially available molecules. Their existence provides scope for using these and other such molecules as hyperpolarising agents in magnetic resonance experiments.

**Acknowledgements**

This work was funded by the EPSRC (EP/F065205/1, EP/H003789/1) and supported by the Oxford e-Research Centre.

# Multiple decoherence-free states in multi-spin systems

H. J. Hogben[1], P. J. Hore[1] and Ilya Kuprov[2]

[1]*Department of Chemistry, University of Oxford, Physical and Theoretical Chemistry Laboratory, South Parks Road, Oxford, OX1 3QZ, UK.*

[2]*Oxford e-Research Centre, University of Oxford, 7 Keble Road, Oxford, OX1 3QG, UK.*

## S.I. EXAMPLE MATLAB CODE

This code should be used as a wrapper calling *Spinach* software. Documentation for *Spinach* can be found in Ref. [1] and the software can be downloaded from http://www.spindynamics.org/Spinach.php.

```matlab
% Outputs
% vR are the eigenvectors of the relaxation matrix
% dR are the eigenvalues  of the relaxation matrix
% vL are the eigenvectors of the total Liouvillian
% dL are the eigenvalues  of the total Liouvillian

% hannah.hogben@chem.ox.ac.uk
% ilya.kuprov@oerc.ox.ac.uk

function [vR,dR,vL,dL]=Long_lived_state_finder()

%% Read the spin system (Gaussian log)
[sys,inter]=g03_to_spinach(g03_parse(['..\molecules\pbenzoquinone.log'])...
,{{'H','1H'}},32);

%% Set the simulation parameters
sys.tols.prox_cutoff=35;
sys.tols.rlx_integration=1e-6;
sys.magnet=1;
sys.hush=1;

% Uncomment the following to zero all chemical shifts.
% inter.zeeman.matrix{1}=speye(3)*(sum(diag(inter.zeeman.matrix{1})/3));
% inter.zeeman.matrix{2}=speye(3)*(sum(diag(inter.zeeman.matrix{2})/3));
% inter.zeeman.matrix{3}=speye(3)*(sum(diag(inter.zeeman.matrix{3})/3));
% inter.zeeman.matrix{4}=speye(3)*(sum(diag(inter.zeeman.matrix{4})/3));

inter.relaxation='redfield';
inter.rlx_keep='full';
inter.tau_c=1e-10;

bas.mode='complete';
spin_system=create(sys,inter);
spin_system=basis(spin_system,bas);

%% Pull out the operators required to make the Liouvillian
R =(r_superop(spin_system));
disp('diagonalising relaxation matrix')
[vR,dR]=eig(full(R));
```

```
L=-1i*h_superop(secularity(spin_system,'keep_all'))+R;
disp('diagonalising total Liouvillian')
[vL,dL]=eig(full(L));

end
```

The console output from *Spinach* when this code is run is as follows:

```
create: LIQUID REGIME, spin system connectivity will be inferred from scalar couplings.
create: 4 spins in the simulation.
create: magnetic induction of 1 Tesla (42.577 MHz proton frequency, -28.025 GHz electron frequency).
create: % summary of non-zero Zeeman interactions (ppm for nuclei, g-tensor for electrons)
---------------------------------------------------------------------------------------------------------------
    Spin  2S+1         Eigenvalues ( A | B | C )           Eigenvectors ( A | B | C )      Isotropic
---------------------------------------------------------------------------------------------------------------
1   1H    2     +2.02224e+00  +7.07755e+00  +9.74721e+00   +1.420e-01  -9.399e-01  +3.104e-01   +6.28233e+00
                                                           +9.233e-01  +1.277e-02  -3.838e-01
                                                           +3.568e-01  +3.411e-01  +8.697e-01

2   1H    2     +2.45950e+00  +7.59105e+00  +9.67926e+00   -2.556e-01  +9.395e-01  +2.280e-01   +6.57660e+00
                                                           -6.913e-01  -1.281e-02  -7.225e-01
                                                           -6.758e-01  -3.423e-01  +6.528e-01

3   1H    2     +2.45950e+00  +7.59105e+00  +9.67926e+00   -2.556e-01  +9.395e-01  +2.280e-01   +6.57660e+00
                                                           -6.913e-01  -1.281e-02  -7.225e-01
                                                           -6.758e-01  -3.423e-01  +6.528e-01

4   1H    2     +2.02224e+00  +7.07755e+00  +9.74721e+00   +1.420e-01  -9.399e-01  +3.104e-01   +6.28233e+00
                                                           +9.233e-01  +1.277e-02  -3.838e-01
                                                           +3.568e-01  +3.411e-01  +8.697e-01
---------------------------------------------------------------------------------------------------------------
create: % summary of non-zero Zeeman interactions (angular frequencies)
---------------------------------------------------------------------------------------------------------------
    Spin  2S+1         Eigenvalues ( A | B | C )           Eigenvectors ( A | B | C )      Isotropic
---------------------------------------------------------------------------------------------------------------
1   1H    2     +5.40994e+02  +1.89340e+03  +2.60760e+03   +1.420e-01  -9.399e-01  +3.104e-01   +1.68066e+03
                                                           +9.233e-01  +1.277e-02  -3.838e-01
                                                           +3.568e-01  +3.411e-01  +8.697e-01

2   1H    2     +6.57970e+02  +2.03077e+03  +2.58942e+03   -2.556e-01  +9.395e-01  +2.280e-01   +1.75939e+03
                                                           -6.913e-01  -1.281e-02  -7.225e-01
                                                           -6.758e-01  -3.423e-01  +6.528e-01

3   1H    2     +6.57970e+02  +2.03077e+03  +2.58942e+03   -2.556e-01  +9.395e-01  +2.280e-01   +1.75939e+03
                                                           -6.913e-01  -1.281e-02  -7.225e-01
                                                           -6.758e-01  -3.423e-01  +6.528e-01

4   1H    2     +5.40994e+02  +1.89340e+03  +2.60760e+03   +1.420e-01  -9.399e-01  +3.104e-01   +1.68066e+03
                                                           +9.233e-01  +1.277e-02  -3.838e-01
                                                           +3.568e-01  +3.411e-01  +8.697e-01
---------------------------------------------------------------------------------------------------------------
create: coordinates (Angstrom)
---------------------------------------
N   Spin      X        Y        Z
---------------------------------------
1   1H     +0.848   -0.312   +2.314
2   1H     +0.352   -2.222   +1.090
3   1H     -0.352   +2.222   -1.090
4   1H     -0.848   +0.312   -2.314
---------------------------------------
create: distance matrix (Angstrom)
---------------------------
N    1     2     3     4
---------------------------
1  0.00  2.32  4.41  4.97
2  2.32  0.00  5.00  4.41
3  4.41  5.00  0.00  2.32
4  4.97  4.41  2.32  0.00
---------------------------
create:          % summary of dipolar interactions
          (angstroms, degrees, angular frequencies)
-----------------------------------------------
L    S     R      Theta     Phi        D/2
-----------------------------------------------
2    1   2.322   121.80   -104.58   +3.0134e+04
3    1   4.410   140.52    115.34   +4.3998e+03
4    1   4.968   158.66    159.81   +3.0770e+03
1    2   2.322    58.20     75.42   +3.0134e+04
3    2   5.000   115.86     98.99   +3.0194e+03
4    2   4.410   140.52    115.34   +4.3998e+03
1    3   4.410    39.48    -64.66   +4.3998e+03
2    3   5.000    64.14    -81.01   +3.0194e+03
4    3   2.322   121.80   -104.58   +3.0134e+04
1    4   4.968    21.34    -20.19   +3.0770e+03
2    4   4.410    39.48    -64.66   +4.3998e+03
3    4   2.322    58.20     75.42   +3.0134e+04
-----------------------------------------------
create: proximity cut-off 35.000 Angstrom.
create: % summary of spin-spin couplings (angular frequencies)
---------------------------------------------------------------------------------------------------------
L    S         Eigenvalues ( A | B | C )              Eigenvectors ( A | B | C )        Isotropic
---------------------------------------------------------------------------------------------------------
1    2    +6.035e+04  +6.035e+04  -1.205e+05     +1.326e-01  -9.678e-01  +2.139e-01    +8.249e+01
                                                 +5.100e-01  +2.517e-01  +8.225e-01
                                                 -8.499e-01  -0.000e+00  +5.270e-01

1    3    +8.813e+03  +8.813e+03  -1.759e+04     +9.038e-01  -3.304e-01  +2.721e-01    +1.383e+01
```

```
                                                        +4.280e-01   +6.976e-01   -5.746e-01
                                                        +0.000e+00   +6.357e-01   +7.719e-01

 1    4   +6.150e+03    +6.150e+03    -1.231e+04        +3.451e-01   -8.742e-01   +3.415e-01   -3.633e+00
                                                        +9.386e-01   +3.214e-01   -1.256e-01
                                                        +0.000e+00   +3.639e-01   +9.315e-01

 2    3   +6.035e+03    +6.035e+03    -1.208e+04        -9.877e-01   +6.816e-02   +1.406e-01   -3.321e+00
                                                        -1.563e-01   -4.307e-01   -8.888e-01
                                                        -0.000e+00   -8.999e-01   +4.361e-01

 2    4   +8.813e+03    +8.813e+03    -1.759e+04        +9.038e-01   -3.304e-01   +2.721e-01   +1.383e+01
                                                        +4.280e-01   +6.976e-01   -5.746e-01
                                                        +0.000e+00   +6.357e-01   +7.719e-01

 3    4   +6.035e+04    +6.035e+04    -1.205e+05        +1.326e-01   -9.678e-01   +2.139e-01   +8.249e+01
                                                        +5.100e-01   +2.517e-01   +8.225e-01
                                                        -8.499e-01   -0.000e+00   +5.270e-01
-----------------------------------------------------------------------------------------------------------
create: connectivity matrix density 100%
create: proximity matrix density 100%

=============================================
=                                           =
=                BASIS SET                  =
=                                           =
=============================================

basis: final basis set summary (L,M quantum numbers in irreducible spherical tensor products).
 N      1        2        3        4
 1    (0,0)    (0,0)    (0,0)    (0,0)
 2    (0,0)    (0,0)    (0,0)    (1,1)
 3    (0,0)    (0,0)    (0,0)    (1,0)
 4    (0,0)    (0,0)    (0,0)    (1,-1)
 5    (0,0)    (0,0)    (1,1)    (0,0)
 6    (0,0)    (0,0)    (1,1)    (1,1)
 7    (0,0)    (0,0)    (1,1)    (1,0)
 8    (0,0)    (0,0)    (1,1)    (1,-1)
 9    (0,0)    (0,0)    (1,0)    (0,0)
10    (0,0)    (0,0)    (1,0)    (1,1)
11    (0,0)    (0,0)    (1,0)    (1,0)
12    (0,0)    (0,0)    (1,0)    (1,-1)
13    (0,0)    (0,0)    (1,-1)   (0,0)
14    (0,0)    (0,0)    (1,-1)   (1,1)
15    (0,0)    (0,0)    (1,-1)   (1,0)
16    (0,0)    (0,0)    (1,-1)   (1,-1)
17    (0,0)    (1,1)    (0,0)    (0,0)
18    (0,0)    (1,1)    (0,0)    (1,1)
19    (0,0)    (1,1)    (0,0)    (1,0)
20    (0,0)    (1,1)    (0,0)    (1,-1)
21    (0,0)    (1,1)    (1,1)    (0,0)
22    (0,0)    (1,1)    (1,1)    (1,1)
23    (0,0)    (1,1)    (1,1)    (1,0)
24    (0,0)    (1,1)    (1,1)    (1,-1)
25    (0,0)    (1,1)    (1,0)    (0,0)
26    (0,0)    (1,1)    (1,0)    (1,1)
27    (0,0)    (1,1)    (1,0)    (1,0)
28    (0,0)    (1,1)    (1,0)    (1,-1)
29    (0,0)    (1,1)    (1,-1)   (0,0)
30    (0,0)    (1,1)    (1,-1)   (1,1)
31    (0,0)    (1,1)    (1,-1)   (1,0)
32    (0,0)    (1,1)    (1,-1)   (1,-1)
33    (0,0)    (1,0)    (0,0)    (0,0)
34    (0,0)    (1,0)    (0,0)    (1,1)
35    (0,0)    (1,0)    (0,0)    (1,0)
36    (0,0)    (1,0)    (0,0)    (1,-1)
37    (0,0)    (1,0)    (1,1)    (0,0)
38    (0,0)    (1,0)    (1,1)    (1,1)
39    (0,0)    (1,0)    (1,1)    (1,0)
40    (0,0)    (1,0)    (1,1)    (1,-1)
41    (0,0)    (1,0)    (1,0)    (0,0)
42    (0,0)    (1,0)    (1,0)    (1,1)
43    (0,0)    (1,0)    (1,0)    (1,0)
44    (0,0)    (1,0)    (1,0)    (1,-1)
45    (0,0)    (1,0)    (1,-1)   (0,0)
46    (0,0)    (1,0)    (1,-1)   (1,1)
47    (0,0)    (1,0)    (1,-1)   (1,0)
48    (0,0)    (1,0)    (1,-1)   (1,-1)
49    (0,0)    (1,-1)   (0,0)    (0,0)
50    (0,0)    (1,-1)   (0,0)    (1,1)
51    (0,0)    (1,-1)   (0,0)    (1,0)
52    (0,0)    (1,-1)   (0,0)    (1,-1)
53    (0,0)    (1,-1)   (1,1)    (0,0)
54    (0,0)    (1,-1)   (1,1)    (1,1)
55    (0,0)    (1,-1)   (1,1)    (1,0)
56    (0,0)    (1,-1)   (1,1)    (1,-1)
57    (0,0)    (1,-1)   (1,0)    (0,0)
58    (0,0)    (1,-1)   (1,0)    (1,1)
59    (0,0)    (1,-1)   (1,0)    (1,0)
60    (0,0)    (1,-1)   (1,0)    (1,-1)
61    (0,0)    (1,-1)   (1,-1)   (0,0)
62    (0,0)    (1,-1)   (1,-1)   (1,1)
63    (0,0)    (1,-1)   (1,-1)   (1,0)
64    (0,0)    (1,-1)   (1,-1)   (1,-1)
65    (1,1)    (0,0)    (0,0)    (0,0)
66    (1,1)    (0,0)    (0,0)    (1,1)
67    (1,1)    (0,0)    (0,0)    (1,0)
68    (1,1)    (0,0)    (0,0)    (1,-1)
69    (1,1)    (0,0)    (1,1)    (0,0)
70    (1,1)    (0,0)    (1,1)    (1,1)
71    (1,1)    (0,0)    (1,1)    (1,0)
72    (1,1)    (0,0)    (1,1)    (1,-1)
```

| | | | | |
|---|---|---|---|---|
| 73 | (1,1) | (0,0) | (1,0) | (0,0) |
| 74 | (1,1) | (0,0) | (1,0) | (1,1) |
| 75 | (1,1) | (0,0) | (1,0) | (1,0) |
| 76 | (1,1) | (0,0) | (1,0) | (1,-1) |
| 77 | (1,1) | (0,0) | (1,-1) | (0,0) |
| 78 | (1,1) | (0,0) | (1,-1) | (1,1) |
| 79 | (1,1) | (0,0) | (1,-1) | (1,0) |
| 80 | (1,1) | (0,0) | (1,-1) | (1,-1) |
| 81 | (1,1) | (1,1) | (0,0) | (0,0) |
| 82 | (1,1) | (1,1) | (0,0) | (1,1) |
| 83 | (1,1) | (1,1) | (0,0) | (1,0) |
| 84 | (1,1) | (1,1) | (0,0) | (1,-1) |
| 85 | (1,1) | (1,1) | (1,1) | (0,0) |
| 86 | (1,1) | (1,1) | (1,1) | (1,1) |
| 87 | (1,1) | (1,1) | (1,1) | (1,0) |
| 88 | (1,1) | (1,1) | (1,1) | (1,-1) |
| 89 | (1,1) | (1,1) | (1,0) | (0,0) |
| 90 | (1,1) | (1,1) | (1,0) | (1,1) |
| 91 | (1,1) | (1,1) | (1,0) | (1,0) |
| 92 | (1,1) | (1,1) | (1,0) | (1,-1) |
| 93 | (1,1) | (1,1) | (1,-1) | (0,0) |
| 94 | (1,1) | (1,1) | (1,-1) | (1,1) |
| 95 | (1,1) | (1,1) | (1,-1) | (1,0) |
| 96 | (1,1) | (1,1) | (1,-1) | (1,-1) |
| 97 | (1,1) | (1,0) | (0,0) | (0,0) |
| 98 | (1,1) | (1,0) | (0,0) | (1,1) |
| 99 | (1,1) | (1,0) | (0,0) | (1,0) |
| 100 | (1,1) | (1,0) | (0,0) | (1,-1) |
| 101 | (1,1) | (1,0) | (1,1) | (0,0) |
| 102 | (1,1) | (1,0) | (1,1) | (1,1) |
| 103 | (1,1) | (1,0) | (1,1) | (1,0) |
| 104 | (1,1) | (1,0) | (1,1) | (1,-1) |
| 105 | (1,1) | (1,0) | (1,0) | (0,0) |
| 106 | (1,1) | (1,0) | (1,0) | (1,1) |
| 107 | (1,1) | (1,0) | (1,0) | (1,0) |
| 108 | (1,1) | (1,0) | (1,0) | (1,-1) |
| 109 | (1,1) | (1,0) | (1,-1) | (0,0) |
| 110 | (1,1) | (1,0) | (1,-1) | (1,1) |
| 111 | (1,1) | (1,0) | (1,-1) | (1,0) |
| 112 | (1,1) | (1,0) | (1,-1) | (1,-1) |
| 113 | (1,1) | (1,-1) | (0,0) | (0,0) |
| 114 | (1,1) | (1,-1) | (0,0) | (1,1) |
| 115 | (1,1) | (1,-1) | (0,0) | (1,0) |
| 116 | (1,1) | (1,-1) | (0,0) | (1,-1) |
| 117 | (1,1) | (1,-1) | (1,1) | (0,0) |
| 118 | (1,1) | (1,-1) | (1,1) | (1,1) |
| 119 | (1,1) | (1,-1) | (1,1) | (1,0) |
| 120 | (1,1) | (1,-1) | (1,1) | (1,-1) |
| 121 | (1,1) | (1,-1) | (1,0) | (0,0) |
| 122 | (1,1) | (1,-1) | (1,0) | (1,1) |
| 123 | (1,1) | (1,-1) | (1,0) | (1,0) |
| 124 | (1,1) | (1,-1) | (1,0) | (1,-1) |
| 125 | (1,1) | (1,-1) | (1,-1) | (0,0) |
| 126 | (1,1) | (1,-1) | (1,-1) | (1,1) |
| 127 | (1,1) | (1,-1) | (1,-1) | (1,0) |
| 128 | (1,1) | (1,-1) | (1,-1) | (1,-1) |
| 129 | (1,0) | (0,0) | (0,0) | (0,0) |
| 130 | (1,0) | (0,0) | (0,0) | (1,1) |
| 131 | (1,0) | (0,0) | (0,0) | (1,0) |
| 132 | (1,0) | (0,0) | (0,0) | (1,-1) |
| 133 | (1,0) | (0,0) | (1,1) | (0,0) |
| 134 | (1,0) | (0,0) | (1,1) | (1,1) |
| 135 | (1,0) | (0,0) | (1,1) | (1,0) |
| 136 | (1,0) | (0,0) | (1,1) | (1,-1) |
| 137 | (1,0) | (0,0) | (1,0) | (0,0) |
| 138 | (1,0) | (0,0) | (1,0) | (1,1) |
| 139 | (1,0) | (0,0) | (1,0) | (1,0) |
| 140 | (1,0) | (0,0) | (1,0) | (1,-1) |
| 141 | (1,0) | (0,0) | (1,-1) | (0,0) |
| 142 | (1,0) | (0,0) | (1,-1) | (1,1) |
| 143 | (1,0) | (0,0) | (1,-1) | (1,0) |
| 144 | (1,0) | (0,0) | (1,-1) | (1,-1) |
| 145 | (1,0) | (1,1) | (0,0) | (0,0) |
| 146 | (1,0) | (1,1) | (0,0) | (1,1) |
| 147 | (1,0) | (1,1) | (0,0) | (1,0) |
| 148 | (1,0) | (1,1) | (0,0) | (1,-1) |
| 149 | (1,0) | (1,1) | (1,1) | (0,0) |
| 150 | (1,0) | (1,1) | (1,1) | (1,1) |
| 151 | (1,0) | (1,1) | (1,1) | (1,0) |
| 152 | (1,0) | (1,1) | (1,1) | (1,-1) |
| 153 | (1,0) | (1,1) | (1,0) | (0,0) |
| 154 | (1,0) | (1,1) | (1,0) | (1,1) |
| 155 | (1,0) | (1,1) | (1,0) | (1,0) |
| 156 | (1,0) | (1,1) | (1,0) | (1,-1) |
| 157 | (1,0) | (1,1) | (1,-1) | (0,0) |
| 158 | (1,0) | (1,1) | (1,-1) | (1,1) |
| 159 | (1,0) | (1,1) | (1,-1) | (1,0) |
| 160 | (1,0) | (1,1) | (1,-1) | (1,-1) |
| 161 | (1,0) | (1,0) | (0,0) | (0,0) |
| 162 | (1,0) | (1,0) | (0,0) | (1,1) |
| 163 | (1,0) | (1,0) | (0,0) | (1,0) |
| 164 | (1,0) | (1,0) | (0,0) | (1,-1) |
| 165 | (1,0) | (1,0) | (1,1) | (0,0) |
| 166 | (1,0) | (1,0) | (1,1) | (1,1) |
| 167 | (1,0) | (1,0) | (1,1) | (1,0) |
| 168 | (1,0) | (1,0) | (1,1) | (1,-1) |
| 169 | (1,0) | (1,0) | (1,0) | (0,0) |
| 170 | (1,0) | (1,0) | (1,0) | (1,1) |
| 171 | (1,0) | (1,0) | (1,0) | (1,0) |
| 172 | (1,0) | (1,0) | (1,0) | (1,-1) |
| 173 | (1,0) | (1,0) | (1,-1) | (0,0) |
| 174 | (1,0) | (1,0) | (1,-1) | (1,1) |

```
175    (1,0)   (1,0)   (1,-1)  (1,0)
176    (1,0)   (1,0)   (1,-1)  (1,-1)
177    (1,0)   (1,-1)  (0,0)   (0,0)
178    (1,0)   (1,-1)  (0,0)   (1,1)
179    (1,0)   (1,-1)  (0,0)   (1,0)
180    (1,0)   (1,-1)  (0,0)   (1,-1)
181    (1,0)   (1,-1)  (1,1)   (0,0)
182    (1,0)   (1,-1)  (1,1)   (1,1)
183    (1,0)   (1,-1)  (1,1)   (1,0)
184    (1,0)   (1,-1)  (1,1)   (1,-1)
185    (1,0)   (1,-1)  (1,0)   (0,0)
186    (1,0)   (1,-1)  (1,0)   (1,1)
187    (1,0)   (1,-1)  (1,0)   (1,0)
188    (1,0)   (1,-1)  (1,0)   (1,-1)
189    (1,0)   (1,-1)  (1,-1)  (0,0)
190    (1,0)   (1,-1)  (1,-1)  (1,1)
191    (1,0)   (1,-1)  (1,-1)  (1,0)
192    (1,0)   (1,-1)  (1,-1)  (1,-1)
193    (1,-1)  (0,0)   (0,0)   (0,0)
194    (1,-1)  (0,0)   (0,0)   (1,1)
195    (1,-1)  (0,0)   (0,0)   (1,0)
196    (1,-1)  (0,0)   (0,0)   (1,-1)
197    (1,-1)  (0,0)   (1,1)   (0,0)
198    (1,-1)  (0,0)   (1,1)   (1,1)
199    (1,-1)  (0,0)   (1,1)   (1,0)
200    (1,-1)  (0,0)   (1,1)   (1,-1)
201    (1,-1)  (0,0)   (1,0)   (0,0)
202    (1,-1)  (0,0)   (1,0)   (1,1)
203    (1,-1)  (0,0)   (1,0)   (1,0)
204    (1,-1)  (0,0)   (1,0)   (1,-1)
205    (1,-1)  (0,0)   (1,-1)  (0,0)
206    (1,-1)  (0,0)   (1,-1)  (1,1)
207    (1,-1)  (0,0)   (1,-1)  (1,0)
208    (1,-1)  (0,0)   (1,-1)  (1,-1)
209    (1,-1)  (1,1)   (0,0)   (0,0)
210    (1,-1)  (1,1)   (0,0)   (1,1)
211    (1,-1)  (1,1)   (0,0)   (1,0)
212    (1,-1)  (1,1)   (0,0)   (1,-1)
213    (1,-1)  (1,1)   (1,1)   (0,0)
214    (1,-1)  (1,1)   (1,1)   (1,1)
215    (1,-1)  (1,1)   (1,1)   (1,0)
216    (1,-1)  (1,1)   (1,1)   (1,-1)
217    (1,-1)  (1,1)   (1,0)   (0,0)
218    (1,-1)  (1,1)   (1,0)   (1,1)
219    (1,-1)  (1,1)   (1,0)   (1,0)
220    (1,-1)  (1,1)   (1,0)   (1,-1)
221    (1,-1)  (1,1)   (1,-1)  (0,0)
222    (1,-1)  (1,1)   (1,-1)  (1,1)
223    (1,-1)  (1,1)   (1,-1)  (1,0)
224    (1,-1)  (1,1)   (1,-1)  (1,-1)
225    (1,-1)  (1,0)   (0,0)   (0,0)
226    (1,-1)  (1,0)   (0,0)   (1,1)
227    (1,-1)  (1,0)   (0,0)   (1,0)
228    (1,-1)  (1,0)   (0,0)   (1,-1)
229    (1,-1)  (1,0)   (1,1)   (0,0)
230    (1,-1)  (1,0)   (1,1)   (1,1)
231    (1,-1)  (1,0)   (1,1)   (1,0)
232    (1,-1)  (1,0)   (1,1)   (1,-1)
233    (1,-1)  (1,0)   (1,0)   (0,0)
234    (1,-1)  (1,0)   (1,0)   (1,1)
235    (1,-1)  (1,0)   (1,0)   (1,0)
236    (1,-1)  (1,0)   (1,0)   (1,-1)
237    (1,-1)  (1,0)   (1,-1)  (0,0)
238    (1,-1)  (1,0)   (1,-1)  (1,1)
239    (1,-1)  (1,0)   (1,-1)  (1,0)
240    (1,-1)  (1,0)   (1,-1)  (1,-1)
241    (1,-1)  (1,-1)  (0,0)   (0,0)
242    (1,-1)  (1,-1)  (0,0)   (1,1)
243    (1,-1)  (1,-1)  (0,0)   (1,0)
244    (1,-1)  (1,-1)  (0,0)   (1,-1)
245    (1,-1)  (1,-1)  (1,1)   (0,0)
246    (1,-1)  (1,-1)  (1,1)   (1,1)
247    (1,-1)  (1,-1)  (1,1)   (1,0)
248    (1,-1)  (1,-1)  (1,1)   (1,-1)
249    (1,-1)  (1,-1)  (1,0)   (0,0)
250    (1,-1)  (1,-1)  (1,0)   (1,1)
251    (1,-1)  (1,-1)  (1,0)   (1,0)
252    (1,-1)  (1,-1)  (1,0)   (1,-1)
253    (1,-1)  (1,-1)  (1,-1)  (0,0)
254    (1,-1)  (1,-1)  (1,-1)  (1,1)
255    (1,-1)  (1,-1)  (1,-1)  (1,0)
256    (1,-1)  (1,-1)  (1,-1)  (1,-1)
basis: state space dimension 256 (100% of the full state space).
basis: no symmetry information available.
 r_superop: computing the lab frame Hamiltonian superoperator...
  secularity: rotating frame assumptions set to "keep_all".
  h_superop: full isotropic Zeeman term for spin 1...
          Lz x 42577748.7985 Hz
  h_superop: full anisotropic Zeeman term for spin 1...
          PHI(-2)  86.6878-41.7643i Hz
          PHI(-1)  19.77937-108.8408i Hz
          PHI( 0)  113.2136 Hz
          PHI( 1) -19.77937-108.8408i Hz
          PHI( 2)  86.6878+41.7643i Hz
  h_superop: full isotropic Zeeman term for spin 2...
          Lz x 42577761.3277 Hz
  h_superop: full anisotropic Zeeman term for spin 2...
          PHI(-2)  24.1755-53.2559i Hz
          PHI(-1) -24.51824-144.0077i Hz
          PHI( 0) -22.9236 Hz
          PHI( 1)  24.51824-144.0077i Hz
```

```
                        PHI( 2)  24.1755+53.2559i Hz
  h_superop: full isotropic Zeeman term for spin 3...
                        Lz x 42577761.3277 Hz
  h_superop: full anisotropic Zeeman term for spin 3...
                        PHI(-2)  24.1755-53.2559i Hz
                        PHI(-1) -24.51824-144.0077i Hz
                        PHI( 0) -22.9236 Hz
                        PHI( 1)  24.51824-144.0077i Hz
                        PHI( 2)  24.1755+53.2559i Hz
  h_superop: full isotropic Zeeman term for spin 4...
                        Lz x 42577748.7985 Hz
  h_superop: full anisotropic Zeeman term for spin 4...
                        PHI(-2)  86.6878-41.7643i Hz
                        PHI(-1)  19.77937-108.8408i Hz
                        PHI( 0)  113.2136 Hz
                        PHI( 1) -19.77937-108.8408i Hz
                        PHI( 2)  86.6878+41.7643i Hz
  h_superop: strong isotropic coupling term for spins 1,2...
                        (LxSx+LySy+LzSz) x 13.128 Hz
  h_superop: strong anisotropic coupling term for spins 1,2...
                        PHI(-2)  9075.63591-5063.51786i Hz
                        PHI(-1) -3243.92797-12472.2762i Hz
                        PHI( 0)  1961.457 Hz
                        PHI( 1)  3243.92797-12472.2762i Hz
                        PHI( 2)  9075.63591+5063.51786i Hz
  h_superop: strong isotropic coupling term for spins 1,3...
                        (LxSx+LySy+LzSz) x 2.2016 Hz
  h_superop: strong anisotropic coupling term for spins 1,3...
                        PHI(-2)  537.9913+656.886i Hz
                        PHI(-1) -882.47589+1863.4214i Hz
                        PHI( 0) -1350.7182 Hz
                        PHI( 1)  882.47589+1863.4214i Hz
                        PHI( 2)  537.9913-656.886i Hz
  h_superop: strong isotropic coupling term for spins 1,4...
                        (LxSx+LySy+LzSz) x -0.57826 Hz
  h_superop: strong anisotropic coupling term for spins 1,4...
                        PHI(-2) -148.195+125.991i Hz
                        PHI(-1) -934.6926+343.624i Hz
                        PHI( 0) -1922.6358 Hz
                        PHI( 1)  934.6926+343.624i Hz
                        PHI( 2) -148.195-125.991i Hz
  h_superop: strong isotropic coupling term for spins 2,3...
                        (LxSx+LySy+LzSz) x -0.52852 Hz
  h_superop: strong anisotropic coupling term for spins 2,3...
                        PHI(-2)  1110.4257+360.44322i Hz
                        PHI(-1) -176.84753+1117.62i Hz
                        PHI( 0)  505.5034 Hz
                        PHI( 1)  176.84753+1117.62i Hz
                        PHI( 2)  1110.4257-360.44322i Hz
  h_superop: strong isotropic coupling term for spins 2,4...
                        (LxSx+LySy+LzSz) x 2.2016 Hz
  h_superop: strong anisotropic coupling term for spins 2,4...
                        PHI(-2)  537.9913+656.886i Hz
                        PHI(-1) -882.47589+1863.4214i Hz
                        PHI( 0) -1350.7182 Hz
                        PHI( 1)  882.47589+1863.4214i Hz
                        PHI( 2)  537.9913-656.886i Hz
  h_superop: strong isotropic coupling term for spins 3,4...
                        (LxSx+LySy+LzSz) x 13.128 Hz
  h_superop: strong anisotropic coupling term for spins 3,4...
                        PHI(-2)  9075.63591-5063.51786i Hz
                        PHI(-1) -3243.92797-12472.2762i Hz
                        PHI( 0)  1961.457 Hz
                        PHI( 1)  3243.92797-12472.2762i Hz
                        PHI( 2)  9075.63591+5063.51786i Hz
 r_superop: largest absolute eigenvalue of the static Hamiltonian superoperator 1069907717.9915
 r_superop: rotational correlation time 1e-10 seconds.
 r_superop: numerical integration time step 6.2798e-11 seconds.
 r_superop: 22 integration steps will be taken.
 propagator: Liouvillian density: 0.027466 %
 propagator: Taylor series converged in 6 iterations.
 propagator: propagator density: 0.39062 %
 r_superop: dynamic frequency shifts have been ignored.
 r_superop: WARNING -- returning full relaxation superoperator, lab frame simulations only.
 r_superop: WARNING -- the spin system will relax to the all-zero state.
diagonalising relaxation matrix
  secularity: rotating frame assumptions set to "keep_all".
  h_superop: full isotropic Zeeman term for spin 1...
                        Lz x 42577748.7985 Hz
  h_superop: full isotropic Zeeman term for spin 2...
                        Lz x 42577761.3277 Hz
  h_superop: full isotropic Zeeman term for spin 3...
                        Lz x 42577761.3277 Hz
  h_superop: full isotropic Zeeman term for spin 4...
                        Lz x 42577748.7985 Hz
  h_superop: strong isotropic coupling term for spins 1,2...
                        (LxSx+LySy+LzSz) x 13.128 Hz
  h_superop: strong isotropic coupling term for spins 1,3...
                        (LxSx+LySy+LzSz) x 2.2016 Hz
  h_superop: strong isotropic coupling term for spins 1,4...
                        (LxSx+LySy+LzSz) x -0.57826 Hz
  h_superop: strong isotropic coupling term for spins 2,3...
                        (LxSx+LySy+LzSz) x -0.52852 Hz
  h_superop: strong isotropic coupling term for spins 2,4...
                        (LxSx+LySy+LzSz) x 2.2016 Hz
  h_superop: strong isotropic coupling term for spins 3,4...
                        (LxSx+LySy+LzSz) x 13.128 Hz
```

## S.II. A FEW MORE COUPLING TOPOLOGIES BASED UPON DIFFERENT DIPOLAR COUPLING STRENGTHS

**TABLE S.II**. Multi-spin systems with both strong ($D \sim -120$ kHz, solid lines) and weak ($D \sim -120$ Hz, dashed lines) dipolar couplings. Nodes represent spin positions. $n$ is the size of the system null space and the nature of the constituent states are listed. $\hat{P}_S^{\{i,j\}}$ is the singlet state of the $i^{th}$ and $j^{th}$ spins and $\sum_i \begin{pmatrix} \hat{1} & \hat{S}_+ & \hat{S}_z & \hat{S}_- \\ \hat{\sigma}_1^i & \hat{\sigma}_2^i & \hat{\sigma}_3^i & \hat{\sigma}_4^i \end{pmatrix}$ is the sum over all permutations of $\hat{1}, \hat{S}_+, \hat{S}_z$ and $\hat{S}_-$ across four spins, where $\hat{\sigma}_n^i$ is the spin state of the $n^{th}$ spin in the $i^{th}$ permutation. Beyond observing that this has, like the singlet state, zero-quantum coherence we offer no physical explanation for this very mixed state. The Kronecker symbol indicates all direct product combinations of the states listed in each set of curly brackets.

| Topology | $n$ | States |
|---|---|---|
| $C_{2h}$ (6-vertex) | 14 | $\hat{\mathbb{1}}^{\{1,2,3,4\}}, \hat{P}_S^{\{1,2\}}, \hat{P}_S^{\{1,2\}}\hat{P}_S^{\{3,4\}}, \hat{P}_S^{\{3,4\}}$ $\hat{P}_S^{\{1,3\}}\hat{P}_S^{\{2,4\}},$ $\hat{P}_S^{\{1,4\}}\hat{P}_S^{\{2,3\}}$ $\sum_i \begin{pmatrix} \hat{\mathbb{1}} & \hat{S}_+ & \hat{S}_z & \hat{S}_- \\ \hat{\sigma}_1^i & \hat{\sigma}_2^i & \hat{\sigma}_3^i & \hat{\sigma}_4^i \end{pmatrix}$ $\hat{P}_S^{\{5,6\}}, \hat{P}_S^{\{1,2\}}\hat{P}_S^{\{5,6\}}, \hat{P}_S^{\{1,2\}}\hat{P}_S^{\{3,4\}}\hat{P}_S^{\{5,6\}}, \hat{P}_S^{\{3,4\}}\hat{P}_S^{\{5,6\}}$ $\hat{P}_S^{\{1,3\}}\hat{P}_S^{\{2,4\}}\hat{P}_S^{\{5,6\}},$ $\hat{P}_S^{\{1,4\}}\hat{P}_S^{\{2,3\}}\hat{P}_S^{\{5,6\}}$ $\sum_i \begin{pmatrix} \hat{\mathbb{1}} & \hat{S}_+ & \hat{S}_z & \hat{S}_- \\ \hat{\sigma}_1^i & \hat{\sigma}_2^i & \hat{\sigma}_3^i & \hat{\sigma}_4^i \end{pmatrix}\hat{P}_S^{\{5,6\}}$ |
| $C_{2h}$ (8-vertex) | 49 | $\left\{\begin{array}{c} \hat{\mathbb{1}}^{\{1,2,3,4\}}, \hat{P}_S^{\{1,2\}}, \hat{P}_S^{\{1,2\}}\hat{P}_S^{\{3,4\}}, \hat{P}_S^{\{3,4\}} \\ \hat{P}_S^{\{1,3\}}\hat{P}_S^{\{2,4\}}, \\ \hat{P}_S^{\{1,4\}}\hat{P}_S^{\{2,3\}} \\ \sum_i \begin{pmatrix} \hat{\mathbb{1}} & \hat{S}_+ & \hat{S}_z & \hat{S}_- \\ \hat{\sigma}_1^i & \hat{\sigma}_2^i & \hat{\sigma}_3^i & \hat{\sigma}_4^i \end{pmatrix} \end{array}\right\}$ $\otimes$ $\left\{\begin{array}{c} \mathbb{1}^{\{5,6,7,8\}}, \hat{P}_S^{\{5,6\}}, \hat{P}_S^{\{5,6\}}\hat{P}_S^{\{7,8\}}, \hat{P}_S^{\{7,8\}} \\ \hat{P}_S^{\{5,7\}}\hat{P}_S^{\{6,8\}}, \\ \hat{P}_S^{\{5,8\}}\hat{P}_S^{\{6,7\}} \\ \sum_i \begin{pmatrix} \hat{\mathbb{1}} & \hat{S}_+ & \hat{S}_z & \hat{S}_- \\ \hat{\sigma}_5^i & \hat{\sigma}_6^i & \hat{\sigma}_7^i & \hat{\sigma}_8^i \end{pmatrix} \end{array}\right\}$ |
| $D_{2h}$ | 240 | too many to list |

## S.III. SIMULATION PARAMETERS FOR THE REAL MOLECULE EXAMPLES

The Gaussian output files [2], for the three example molecules can be found in the *Spinach* repository in the molecules folder, (…/spinach/exp/molecules). The calculations were carried out for the fully protonated molecules and hydrogens that were required to be deuterium were renamed as He in the output file simply to prevent the file parser recognizing them. Deuterium nuclei were assumed to have no dipolar couplings. The important parameters are summarised below.

General parameters

- Static field, $B_0$ = 0.1 T
- Isotropic rotational correlation time, $\tau_c = 1 \times 10^{-10}$ s

Example 1 – Bicyclopropylidene [3]

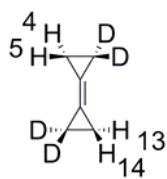

Figure 1. Bicyclopropylidene. Atom numbering is as in the Gaussian output file

**TABLE S.III**. Atomic parameters for bicyclopropylidene calculated with Gaussian03 (using B3LYP functional and EPR-II basis. NMR properties were computed using the GIAO method). Atom numbering is as in the Gaussian output file.

| Atom | Atomic coordinates / Å | | | Chemical shift tensor / ppm | | |
|---|---|---|---|---|---|---|
| 4 | 1.878000 | -0.332000 | 1.779200 | XX=31.9666  XY=-5.0811  XZ= 5.3750 | YX=-0.8605  YY=37.6188  YZ=-7.1735 | ZX= 0.6082  ZY=-6.4210  ZZ=36.9629 |
| 5 | 2.430600 | 0.651700 | 0.690600 | XX=41.2897  XY= 6.0678  XZ=-7.5668 | YX= 1.4425  YY=31.6434  YZ=-1.6927 | ZX=-3.2160  ZY=-2.2064  ZZ=33.3079 |
| 13 | -0.604000 | -2.226700 | -1.186600 | XX=30.4288  XY=-0.6744  XZ=-0.1147 | YX= 2.6108  YY=44.6410  YZ=-5.8230 | ZX=-0.2093  ZY=-0.3937  ZZ=31.3936 |
| 14 | -0.051500 | -1.243000 | -2.274700 | XX=34.0984  XY= 1.6543  XZ=-6.5945 | YX=-1.2143  YY=30.6556  YZ= 2.7798 | ZX=-6.1301  ZY=-2.8968  ZZ=41.9037 |

**TABLE S.IV**. Scalar coupling parameters for bicyclopropylidene, in Hz, calculated by Gaussian03 (using B3LYP/EPR-II with GIAO). Atom numbering is as in the Gaussian output file.

| | 4 | 5 | 13 |
|---|---|---|---|
| 5   | -16.5886 |         |          |
| 13  | 2.14001  | 2.60914 |          |
| 14  | 2.56393  | 2.15407 | -16.6225 |

Example 2 – Naphthalenetetrone [4]

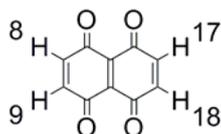

Figure 2. Naphthalenetetrone. Atom numbering is as in the Gaussian output file

**TABLE S.V**. Atomic parameters for naphthalenetetrone calculated by Gaussian03 (using B3LYP functional and EPR-II basis. NMR properties were computed using the GIAO method). Atom numbering is as in the Gaussian output file.

| Atom | Atomic coordinates / Å | | | Chemical shift tensor / ppm | | |
|------|---------|----------|----------|-----------------|-----------------|-----------------|
| 8    | 1.033300 | -3.199200 | -1.104100 | XX=31.7033<br>XY= 3.5559<br>XZ= 0.1729 | YX= 3.6987<br>YY=23.5725<br>YZ=-1.2690 | ZX=-0.5024<br>ZY=-1.9032<br>ZZ=28.5119 |
| 9    | 2.756400 | -1.665700 | -1.491800 | XX=25.3357<br>XY= 4.3285<br>XZ= 0.8655 | YX= 4.4757<br>YY=30.2041<br>YZ=-1.3147 | ZX= 2.3360<br>ZY=-1.4728<br>ZZ=27.5992 |
| 17   | -2.756400 | 1.665700 | 1.491800 | XX=25.3357<br>XY= 4.3285<br>XZ= 0.8655 | YX= 4.4757<br>YY=30.2041<br>YZ=-1.3147 | ZX= 2.3360<br>ZY=-1.4728<br>ZZ=27.5992 |
| 18   | -1.033300 | 3.199200 | 1.104100 | XX=31.7033<br>XY= 3.5559<br>XZ= 0.1728 | YX= 3.6987<br>YY=23.5725<br>YZ=-1.2690 | ZX=-0.5024<br>ZY=-1.9032<br>ZZ=28.5119 |

**TABLE S.VI**. Scalar coupling parameters for naphthalenetetrone, in Hz, calculated by Gaussian03 (using B3LYP/EPR-II with GIAO). Atom numbering is as in the Gaussian output file.

|    | 8         | 9         | 17      |
|----|-----------|-----------|---------|
| 9  | 9.25793   |           |         |
| 17 | -0.293323 | -0.383734 |         |
| 18 | -0.401759 | -0.293323 | 9.25793 |

Example 3 – p-Benzoquinone [5]

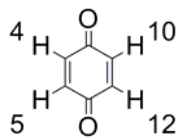

Figure 3. p-Benzoquinone. Atom numbering is as in the Gaussian output file

**TABLE S.VII**. Atomic parameters for p-benzoquinone calculated by Gaussian03 (using B3LYP functional and EPR-II basis. NMR properties were computed using the GIAO method). Atom numbering is as in the Gaussian output file..

| Atom | Atomic coordinates / Å | | | Chemical shift tensor / ppm | | |
|---|---|---|---|---|---|---|
| 4 | 0.848400 | -0.311900 | 2.313900 | XX=24.7672<br>XY= 0.8491<br>XZ=-0.4283 | YX= 1.1127<br>YY=28.8392<br>YZ= 2.9854 | ZX=-0.5008<br>ZY= 2.1272<br>ZZ=23.5466 |
| 5 | 0.351600 | -2.222000 | 1.090200 | XX=24.6358<br>XY= 1.4291<br>XZ= 0.5845 | YX= 1.0725<br>YY=25.7714<br>YZ= 2.9719 | ZX= 0.5672<br>ZY= 3.7926<br>ZZ=25.8630 |
| 10 | -0.351600 | 2.222000 | -1.090200 | XX=24.6358<br>XY= 1.4291<br>XZ= 0.5845 | YX= 1.0725<br>YY=25.7714<br>YZ= 2.9719 | ZX= 0.5672<br>ZY= 3.7926<br>ZZ=25.8630 |
| 12 | -0.848400 | 0.311900 | -2.313900 | XX=24.7672<br>XY= 0.8491<br>XZ=-0.4283 | YX= 1.1127<br>YY=28.8392<br>YZ= 2.9854 | ZX=-0.5008<br>ZY= 2.1272<br>ZZ=23.5466 |

**TABLE S.VIII**. Scalar coupling parameters for p-benzoquinone, in Hz, calculated by Gaussian03 (using B3LYP/EPR-II with GIAO). Atom numbering is as in the Gaussian output file.

|    | 4 | 5 | 10 |
|---|---|---|---|
| 5  | 13.1280 | | |
| 10 | 2.20162 | -0.528524 | |
| 12 | -0.578260 | 2.20162 | 13.1280 |